\documentclass[twocolumn,prb,showpacs,amsmath,amssymb]{revtex4}

\usepackage{graphicx}% Include figure files
\usepackage{dcolumn}% Align table columns on decimal point
\usepackage{bm}% bold math

\begin{document}

\title{Superconductivity in a model of two Hubbard chains coupled with \\
ferromagnetic exchange interaction}

\author{T. Shirakawa}
\affiliation{
Graduate School of Science and Technology, 
Chiba University, Chiba 263-8522, Japan}

\author{S. Nishimoto}
\affiliation{
Leibniz-Institut f\"ur Festk\"orper- und Werkstoffforschung Dresden, 
P.O. Box 270116, 01171 Dresden, Germany
}

\author{Y. Ohta}
\affiliation{
Department of Physics, Chiba University, Chiba 263-8522, Japan
}

\date{\today}

\begin{abstract}
We study the ground-state properties of the double-chain Hubbard model 
coupled with ferromagnetic exchange interaction by using the weak-coupling 
theory, density-matrix renormalization group technique, and Lanczos 
exact-diagonalization method. 
We determine the ground-state phase diagram in the parameter space of the 
ferromagnetic exchange interaction and band filling.  We find that, 
in high electron density regime, the spin gap opens and the spin-singlet 
$d_{xy}$-wave-like pairing correlation is most dominant, whereas in low 
electron density regime, the fully-polarized ferromagnetic state is 
stabilized where the spin-triplet $p_{y}$-wave-like pairing correlation 
is most dominant.  
\end{abstract}

\pacs{71.10.Fd, 71.10.Hf, 74.20.Mn, 74.20.Rp}

\maketitle

\section{Introduction}
Ferromagnetism in itinerant electron systems has increasingly been understood 
since the Hubbard model was introduced in 1963.~\cite{Hubbard63,Kanamori63}  
It is known that subtle interplay between quantum many-body effects and 
spin-independent Coulomb interactions plays a crucial role in generating 
ferromagnetic orders in some solids;~\cite{Mielke93} a variety of origins 
such as Nagaoka-Thouless mechanism,~\cite{Nagaoka66,Thouless65} 
flat-band,~\cite{Mielke91,Tasaki92} 
orbital degeneracy,~\cite{Zener51,VanVleck53,Slater53} 
three-site ring exchange interaction,~\cite{Penc96,Fazekas99} etc., have 
so far been proposed.  
In addition to ferromagnetism, (possibly) spin-triplet superconductivity 
was recently discovered in the metallic ferromagnets UGe$_2$,~\cite{Saxena00} 
URhGe,~\cite{Aoki01} and ZrZn$_2$.~\cite{Pfleiderer}  
Consequently, the relation between superconductivity and ferromagnetism has 
been of special interest in the field of strongly correlated systems.  
From the theoretical point of view, the occurrence of superconductivity 
in ferromagnetic materials is naturally explained by the formation of Cooper 
pairs with parallel spins, namely, spin-triplet pairs.~\cite{Fay80}  

Among the origins of ferromagnetism mentioned above, only the three-site ring 
exchange interaction acts on a couple of electrons.  It yields a ferromagnetic 
spin correlation, which in turn produces an attractive effect between them.  
One may easily imagine that a spin-triplet superconductivity is realized if 
the attractive interaction between electrons can survive against the other 
effects.  
Recently, we have confirmed that this mechanism actually works in a fairly simple 
correlated electron system; the system consists of two Hubbard chains coupled 
with zigzag bonds and has a unique structure of hopping 
integrals.~\cite{Ohta05,Nishimoto08a,Nishimoto08b}  
In this model, the spin-triplet pairing of electrons occurs between the 
inter-chain neighboring sites.  If the ferromagnetic correlation between the 
two chains are essential for the spin-triplet superconductivity, we may be 
allowed to mimic our original model by a double-chain Hubbard model coupled 
with ferromagnetic exchange interaction.  This new model is much easier to 
analyze than the original one due to the reduction of quantum fluctuations.  
Therefore, the introduction of this model will enable us to investigate the 
spin-triplet superconductivity in more detail for wide range of parameters.  

In this paper, we thus study the double-chain Hubbard model coupled with 
ferromagnetic exchange interaction.  
We use the weak-coupling bosonization/renormalization group (RG) 
analyses,~\cite{Solyom79} the density-matrix renormalization group (DMRG) 
technique, and the Lanczos exact-diagonalization method.~\cite{Lanczos56} 
We thereby determine the ground-state phase diagram: we find that, in the 
high electron density regions, the spin gap opens and the $d_xy$-wave-like 
pairing occurs, while in the low electron density regions, the system is 
in the metallic state with full spin polarization, where the $p_y$-wave-like 
spin-triplet pairing correlation becomes the most dominant.  
We note that this model can be regarded as a single-chain model with two 
degenerate orbitals if we assume the ferromagnetic exchange interaction to 
be identified with the intra-atomic Hund's rule coupling.  
This single-chain model has so far been studied both 
analytically~\cite{Fujimoto95,Khveshchenko94,Shelton96} and 
numerically~\cite{Ammon00,Sakamoto02}.  
It has been proposed that the Haldane gap state is realized at half filling.  
It has also been pointed out that the system remains gapful for low hole 
doping regions.~\cite{Nagaosa96}

Our paper is organized as follows.  In Sec.~\ref{sec:level1}, we define 
the double-chain Hubbard model coupled with ferromagnetic exchange 
interaction.  In Sec.~\ref{sec:weakcoupling}, we analyze the model using 
the weak-coupling theory and derive the pairing order parameter.  
In Sec.~\ref{sec:numerical}, we calculate several quantities with the 
DMRG and Lanczos methods and present the calculated results.  
We also check if the results are consistent with the weak-coupling 
results.  Section \ref{sec:summary} contains summary and conclusions.  

\section{Model}
\label{sec:level1}

\begin{figure}
\includegraphics[width=0.6\hsize]{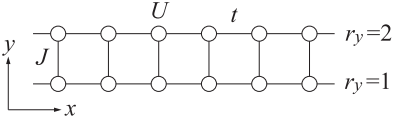}
\caption{\label{fig:ladder} Schematic representation 
of the double-chain Hubbard model coupled with ferromagnetic 
exchange interaction. No hopping process is allowed between 
the two chains and no exchange interactions exist along the chains.
}
\end{figure}

We study the double-chain Hubbard model coupled with ferromagnetic 
exchange interaction (see Fig. \ref{fig:ladder}), 
the Hamiltonian of which is defined by
\begin{eqnarray}
H = && t \sum_{r_x, r_y, \sigma} 
\left( c_{r_x,r_y,\sigma}^{\dagger}c_{r_x+1,r_y,\sigma} 
+ {\rm H.c.} \right) \nonumber\\
&& + U \sum_{r_x, r_y} n_{r_x,r_y,\uparrow}n_{r_x,r_y,\downarrow} 
\nonumber \\
&& - J \sum_{r_x} {\bf S}_{r_x,1}\cdot {\bf S}_{r_x,2}
\label{eq:ham}
\end{eqnarray}
where $c_{r_x,r_y,\sigma}^{\dagger}$ ($c_{r_x,r_y,\sigma}$) is 
the creation (annihilation) operator of an electron with spin 
$\sigma$ ($=\uparrow,\downarrow$) at site $r_x$ on leg $r_y$ ($=1,2$), 
$n_{r_x,r_y,\sigma}=c_{r_x,r_y,\sigma}^{\dagger}c_{r_x,r_y,\sigma}$ 
is the density operator, and ${\bf S}_{r_x,r_y}$ is the 
spin-$\frac{1}{2}$ operator. 
$t$ is the nearest-neighbor hopping integral along the chain, 
$U$ is the onsite Coulomb interaction, and $J (>0)$ is the 
ferromagnetic exchange interaction between two sites on each rung.  
Note that no hopping process is allowed between the two chains 
and thus the two bands are degenerate.

\section{Weak-coupling theory}
\label{sec:weakcoupling}

We first consider the weak-coupling regime where only the 
low-energy excitations near the Fermi points are crucial.  
Thus far, the weak-coupling theory has been well-constructed 
for two-band models.~\cite{Varma85,Nersesyan91,Fabrizio93,Khveshchenko94,Schulz96,Balents96,Orignac97,Tsuchiizu01,Fjarestad02}  
We further develop the approach to analyze the Hamiltonian 
(\ref{eq:ham}).
Assuming a linearization of the dispersion relations in the 
vicinity of the Fermi level, we introduce the field operators 
of right- and left-going electrons as 
\begin{equation}
\psi_{p,\sigma,\pm}(x) = \frac{1}{\sqrt{L}} 
\sum_{k_x} e^{ikx} c_{p,\sigma}^\pm(k)
\label{eq:fo}
\end{equation}
where $c_{p,\sigma}^{\pm}$ is the Fourier transform of combined 
operator $(c_{r_x,1,\sigma} \pm c_{r_x,1,\sigma})/\sqrt{2}$ for 
the right-going ($p=+$) and left-going ($p=-$) electrons.  
$L$ is the chain length where the lattice spacing is set to be unity.  
Using these field operators, the Hamiltonian can be rewritten 
as a sum of the linearized kinetic energy and interaction terms.  
We thus obtain
\begin{eqnarray}
H = && \int dx H_0 + \int dx H_I \nonumber \\
H_0 = && v_{\text F} \sum_{p,\sigma, \zeta} 
\psi_{p,\sigma,\zeta}^{\dagger}
\left( -ip \frac{d}{dx}  \right) \psi_{p,\sigma,\zeta} 
\nonumber \\
H_I = && \frac{1}{4} \sum_{p,\sigma}\sum_{\zeta}~\hspace*{-1.0mm}^{\prime}
g_{1\perp}^{\epsilon,\bar{\epsilon}} 
\psi_{p,\sigma,\zeta_1}^{\dagger} 
\psi_{-p,-\sigma,\zeta_2}^{\dagger} 
\psi_{p,-\sigma,\zeta_4} 
\psi_{-p,\sigma,\zeta_3} \nonumber \\
&& \frac{1}{4} \sum_{p,\sigma}\sum_{\zeta}~\hspace*{-1.0mm}^{\prime} 
g_{2\perp}^{\epsilon,\bar{\epsilon}} 
\psi_{p,\sigma,\zeta_1}^{\dagger} 
\psi_{-p,-\sigma,\zeta_2}^{\dagger} 
\psi_{-p,-\sigma,\zeta_4}
\psi_{p,\sigma,\zeta_3} \nonumber \\
&& \frac{1}{4} \sum_{p,\sigma}\sum_{\zeta}~\hspace*{-1.0mm}^{\prime} 
g_{\parallel}^{\epsilon,\bar{\epsilon}} 
\psi_{p,\sigma,\zeta_1}^{\dagger} 
\psi_{-p,\sigma,\zeta_2}^{\dagger} 
\psi_{p,\sigma,\zeta_4}
\psi_{-p,\sigma,\zeta_3}.
\label{eq:ham2}
\end{eqnarray}
where $\epsilon = \zeta_1 \zeta_3$ and $\bar{\epsilon}=\zeta_1\zeta_2$.  
The primed summation over $\zeta_i (i = 1,2,3,4)$ is restricted by 
a relation $\zeta_1 \zeta_2 \zeta_3 \zeta_4 = +1$, which comes from 
the momentum conservation condition in the transverse components.  
The coupling constants $g_{1\perp}^{\epsilon,\bar{\epsilon}}$ and 
$g_{2\perp}^{\epsilon,\bar{\epsilon}}$ are related to the original 
parameters in the Hamiltonian (\ref{eq:ham}):
\begin{eqnarray}
&& g_{1\perp}^{++} = U, \quad g_{1\perp}^{+-} = - \frac{J}{4}, 
\quad g_{1\perp}^{--} = - \frac{J}{2}, \nonumber \\
&& 
g_{2\perp}^{++} = U, \quad g_{2\perp}^{+-} = -\frac{J}{4}, 
\quad g_{2\perp}^{--} = - \frac{J}{2}. 
\end{eqnarray}
For the SU(2) symmetric case, we can choose 
\begin{eqnarray}
g_{\parallel}^{++} && = g_{1\perp}^{++}-g_{2\perp}^{++} \\
g_{\parallel}^{-+} && = g_{1\perp}^{-+}-g_{2\perp}^{-+} \\
g_{\parallel}^{+-} && = g_{1\perp}^{+-}-g_{2\perp}^{--} \\
g_{\parallel}^{--} && = g_{1\perp}^{--}-g_{2\perp}^{+-}.
\end{eqnarray}
In this Section III, we consider the case away from half filling.  
Hence, the Umklapp term $g_3$ gives no contribution and the Fermi 
velocity renormalization due to the forward-scattering term $g_4$ 
may be neglected. 

\subsection{Bosonization}

Using the Abelian bosonization method,~\cite{Emery79} we introduce 
eight chiral bosonic fields $\phi_{\mu r}^p$ where $\mu$ refers 
to the charge ($\rho$) and spin ($\sigma$) sectors; meanwhile, $r$ 
refers to the even ($+$) and odd ($-$) sectors.  
The bosonic fields satisfy the commutation relations 
$[ \phi_{\mu r}^{\pm}(x), \phi_{\mu^{\prime} r^{\prime}}^{\pm}(x^{\prime}) ] 
= \pm i (\pi/4) \mathrm{sgn} (x-x^{\prime}) \delta_{\mu,\mu^{\prime}} \delta_{r,r^{\prime}}$ 
and $[ \phi_{\mu r}^+(x), \phi_{\mu^{\prime} r^{\prime}}^-(x^{\prime}) ] = 
i (\pi/4)\delta_{\mu,\mu^{\prime}} \delta_{r,r^{\prime}}$.  
We then define a new set of chiral bosonic fields as
\begin{eqnarray}
\phi_{p,s,\zeta} = 
\phi_{\rho +}^p + \zeta \phi_{\rho -}^p
+s \phi_{\sigma +}^p + s \zeta \phi_{\sigma -}^p
\end{eqnarray}
where $p=\pm$, $s=\pm$, and $\zeta=\pm$.  
The chiral bosons obey the commutation relations 
$[ phi_{p,s,\zeta}(x), phi_{p,s^{\prime},\zeta^{\prime}}(x^{\prime}) ] 
= \pm i (\pi/4) \mathrm{sgn}(x-x^{\prime}) \delta_{p,p^{\prime}} \delta_{s,s^{\prime}}$ 
and $[ \phi_{+,s,\zeta}(x),  \phi_{-,s^{\prime},\zeta^{\prime}}(x^{\prime}) ] = 
i (\pi/4)\delta_{s,s^{\prime}} \delta_{\zeta,\zeta^{\prime}}$.  
The field operators of the right-moving and left-moving electrons 
(\ref{eq:fo}) are then written as 
\begin{eqnarray}
\psi_{p,\sigma,\zeta} = \frac{\eta_{\sigma,\zeta}}{\sqrt{2\pi a}} 
\exp \left( ipk_{\text F} x + ip \phi_{p,s,\zeta}  \right)
\end{eqnarray}
where $s= +(-)$ for $\sigma = \uparrow(\downarrow)$.  
The Majorana fermions $\eta_{\sigma,\zeta}$, known as Klein factors, 
are introduced to ensure the proper anticommutation relations between 
fermion fields with different band and spin indices.  They obey 
$\left\{ \eta_{\sigma,\zeta},\eta_{\sigma^{\prime},\zeta^{\prime}} \right\}
= 2 \delta_{\sigma,\sigma^{\prime}} \delta_{\zeta,\zeta^{\prime}}$.  
It is generally more convenient to trade the chiral boson fields 
pairwise for a conventional bosonic phase field $\phi$ and its
dual field $\theta$, so that we also introduce the bosonic fields 
given by 
\begin{eqnarray}
&& \phi_{\mu,r} = \phi_{\mu,r}^+ + \phi_{\mu,r}^- \\
&& \theta_{\mu,r} = \phi_{\mu,r}^+ - \phi_{\mu,r}^- ,
\end{eqnarray}
where the operator $\Pi_{\mu r}(x) = \partial_x \theta_{\mu r}/\pi$ is 
a canonical conjugate variable to $\phi_{\mu r}$, which satisfies 
$\left[\phi_{\mu r}(x), \Pi_{\mu r}(x^{\prime}) \right]
=i\delta (x-x^{\prime} ) \delta_{\mu,\mu^{\prime}} 
\delta_{r,r^{\prime}}$.  

Now, we can rewrite the noninteracting term of the Hamiltonian 
(\ref{eq:ham2}) as 
\begin{eqnarray}
H_0 = \frac{v_{\text F}}{2\pi} \sum_{\mu = \rho,\sigma} 
\sum_{r = \pm} \left[ 
\left(\partial_x \theta_{\mu,r} \right)^2 +
\left(\partial_x \phi_{\mu,r} \right)^2
\right]. 
\label{eq:kinetic}
\end{eqnarray}
On the other hand, the interacting term is more complicated, 
containing many products of the Klein factors such as 
$\hat{\Gamma} \equiv \eta_{\uparrow,+}\eta_{\downarrow,+}\eta_{\uparrow,-}\eta_{\uparrow,-}$, 
$\hat{h_{\sigma}} \equiv \eta_{\sigma,+}\eta_{\sigma,-}$, and 
$\hat{h_{\zeta}}^{\prime} \equiv \eta_{\uparrow,\zeta}\eta_{\downarrow,\zeta}$. 
However, it is known that their eigenvalues are $\Gamma = \pm 1$, $h_{\sigma} = \pm i$, 
and $h_{\zeta}^{\prime} = \pm i$.~\cite{Fjarestad02}  
Thus, if we adopt the following convention 
$\Gamma = +1$, $h_{\sigma} = i$, $h_{\zeta}^{\prime} = i \zeta$, 
the interacting term is reduced to
\begin{eqnarray}
H_I &&= \sum_{\mu = \rho,\sigma} \sum_{r = \pm} 
\frac{g_{\mu,r}}{2\pi^2} 
\partial_x \phi_{\mu,r}^+ \partial_x \phi_{\mu,r}^-
+ \nonumber \\
&& \frac{1}{2\pi^2} \left[ 
\left(g_{1\perp}^{+-}-g_{2\perp}^{--}\right) 
\cos 2 \phi_{\rho -} \cos 2 \phi_{\sigma -} \right. \nonumber \\
&& 
+ \left(g_{1\perp}^{-+}-g_{2\perp}^{-+}\right) 
\cos 2 \theta_{\rho -} \cos 2 \theta_{\sigma -} \nonumber \\
&& 
+ g_{1\perp}^{++}
\cos 2 \phi_{\sigma +} \cos 2 \phi_{\sigma -} 
+g_{1\perp}^{+-}
\cos 2 \phi_{\rho -} \cos 2 \phi_{\sigma +} \nonumber \\
&& 
-g_{1\perp}^{-+}
\cos 2 \theta_{\rho -} \cos 2 \phi_{\sigma +} 
+g_{1\perp}^{--}
\cos 2 \phi_{\sigma +} \cos 2 \theta_{\sigma -} \nonumber \\
&& 
\left.
-g_{2\perp}^{-+}
\cos 2 \theta_{\rho -} \cos 2 \phi_{\sigma -} 
+g_{2\perp}^{--}
\cos 2 \phi_{\rho -} \cos 2 \theta_{\sigma -} \right] \nonumber \\
\label{eq:hamint}
\end{eqnarray}
with 
\begin{eqnarray}
&& g_{\rho + } = - g_{1\perp}^{++} + 2g_{2\perp}^{++} 
-g_{1\perp}^{--} + g_{2\perp}^{+-} \\
&& g_{\rho - } = - g_{1\perp}^{++} + 2g_{2\perp}^{++} 
+ g_{1\perp}^{--} - g_{2\perp}^{+-} \\
&& g_{\sigma + } = - g_{1\perp}^{++} - 2g_{1\perp}^{--} \\
&& g_{\sigma - } = - g_{1\perp}^{++} + 2g_{1\perp}^{--}. 
\end{eqnarray}
If we use the notation with $g_{\mu,r}$, $H_0$ can also be written as
\begin{eqnarray}
H_0^{\prime} =  \sum_{\mu,r} 
\frac{u_{\mu,r}}{2\pi}
\left[ 
K_{\mu,r}\left(\partial_x \theta_{\mu,r}\right)^2 + 
K_{\mu,r}^{-1} \left(\partial_x \phi_{\mu,r}\right)^2
\right]
\end{eqnarray}
with the critical exponents
\begin{eqnarray}
K_{\mu,r} = 
\sqrt{\frac{2\pi v_{\text F}-g_{\mu,r}}{2\pi v_{\text F}+g_{\mu,r}}}
\end{eqnarray}
and the renormalized Fermi velocity
\begin{eqnarray}
u_{\mu,r} = v_{\text F} 
\sqrt{1-\left(\frac{g_{\mu,r}}{2\pi v_{\text F}} \right)^2}. 
\end{eqnarray}

\subsection{Renormalization-group analysis}

By treating the interaction perturbatively, we derive the renormalization 
group (RG) equations in the one-loop level as follows:
\begin{eqnarray}
\frac{\partial g_{1\perp}^{++}}{\partial S} &=&
- 2 \left( g_{1\perp}^{++} \right)^2
- 2g_{1\perp}^{-+} g_{2\perp}^{-+} \nonumber \\
&&- 2 \left( g_{1\perp}^{+-} \right)^2
+ 2g_{1\perp}^{+-} g_{2\perp}^{--} 
\label{RGeqF}
\end{eqnarray}
\begin{eqnarray}
\frac{\partial g_{1\perp}^{-+}}{\partial S} &=&
- 2g_{1\perp}^{-+} g_{2\perp}^{++} 
- 2g_{2\perp}^{-+} g_{1\perp}^{++} 
- 4g_{1\perp}^{-+} g_{1\perp}^{--} \nonumber \\ 
&&+ 2 \left( g_{1\perp}^{-+}g_{2\perp}^{+-} + g_{1\perp}^{--}g_{2\perp}^{-+} \right)
\end{eqnarray}
\begin{eqnarray}
\frac{\partial g_{--}^{1\perp}}{\partial S} &=&
- 2 \left( g_{1\perp}^{-+} \right)^2 - 2 \left( g_{1\perp}^{--} \right)^2 
\nonumber \\
&&+ 2g_{1\perp}^{-+}g_{2\perp}^{-+} 
- 2g_{1\perp}^{+-}g_{2\perp}^{--} 
\end{eqnarray}
\begin{eqnarray}
\frac{\partial g_{1\perp}^{+-}}{\partial S} &=&
- 2 \left(2 g_{1\perp}^{+-} -g_{2\perp}^{--} \right) 
g_{1\perp}^{++}
+ 2g_{1\perp}^{+-}g_{2\perp}^{++} \nonumber \\
&&- 2g_{1\perp}^{--}g_{2\perp}^{--} 
- 2g_{1\perp}^{+-}g_{2\perp}^{+-} 
\end{eqnarray}
\begin{eqnarray}
\frac{\partial g_{2\perp}^{++}}{\partial S} =
- \left( g_{1\perp}^{++} \right)^2 
- \left( g_{1\perp}^{-+} \right)^2 
- \left( g_{2\perp}^{-+} \right)^2 
+ \left( g_{2\perp}^{--} \right)^2 
\end{eqnarray}

\begin{eqnarray}
\frac{\partial g_{2\perp}^{-+}}{\partial S} =
- 2g_{1\perp}^{++}g_{1\perp}^{-+} 
- 2g_{2\perp}^{++}g_{2\perp}^{-+} 
+ 2g_{2\perp}^{-+}g_{2\perp}^{+-} 
\end{eqnarray}

\begin{eqnarray}
\frac{\partial g_{2\perp}^{--}}{\partial S} =
- 2 g_{1\perp}^{--}g_{1\perp}^{+-} 
+ 2 g_{2\perp}^{++}g_{2\perp}^{--} 
- 2 g_{2\perp}^{--}g_{2\perp}^{+-} 
\end{eqnarray}

\begin{eqnarray}
\frac{\partial g_{2\perp}^{+-}}{\partial S} &=&
- \left( g_{1\perp}^{--} \right)^2
- \left( g_{1\perp}^{+-} \right)^2 \nonumber \\
&&+ \left( g_{2\perp}^{-+} \right)^2
- \left( g_{2\perp}^{--} \right)^2
\label{RGeqL}
\end{eqnarray}
where $S=\ln l/(\pi v_F)$ is the RG time and $l$ is the scaling quantity.  
We note that the couplings $g_{1\perp}^{--}$ and $g_{2\perp}^{+-}$ play 
important roles here.  
In general, those couplings are irrelevant when the hopping processes 
between two chains are 
relevant~\cite{Fabrizio93,Khveshchenko94,Schulz96,Balents96,Tsuchiizu01}.  
However, there is no transverse hopping term in our model, so that we have 
to start from the picture of degenerate bands.

By solving the RG equations (\ref{RGeqF})-(\ref{RGeqL}), we obtain 
the relations 
\begin{eqnarray}
g_{1\perp}^{+-}, g_{2\perp}^{--}, -g_{1\perp}^{--}, 
-g_{2\perp}^{+-} \gg g_{2\perp}^{++} \gg g_{1\perp}^{++} \gg 0
\end{eqnarray}
and $g_{-+}^{1\perp} = g_{2\perp}^{-+} = 0$ for $U \gtrsim J$.  
In this case, the parameters $K_{\rho -}$, $K_{\sigma +}$, and 
$K_{\sigma -}$ are scaled as $K_{\rho -} \to 0$, $K_{\sigma +}\to 0$, 
and $K_{\sigma -} \to \infty$.  
We therefore can simplify the second term of the phase Hamiltonian 
(\ref{eq:hamint}) 
as
\begin{eqnarray}
&& 
\left(g_{1\perp}^{+-}-g_{2\perp}^{--}\right) 
\cos 2 \phi_{\rho -} \cos 2 \phi_{\sigma -} \nonumber \\
&& 
+g_{1\perp}^{++}
\cos 2 \phi_{\sigma +} \cos 2 \phi_{\sigma -} 
+g_{1\perp}^{+-}
\cos 2 \phi_{\rho -} \cos 2 \phi_{\sigma +} \nonumber \\
&& 
+g_{1\perp}^{--}
\cos 2 \phi_{\sigma +} \cos 2 \theta_{\sigma -} 
+g_{2\perp}^{--}
\cos 2 \phi_{\rho -} \cos 2 \theta_{\sigma -}. \nonumber \\
\end{eqnarray}
Note that the phase variables $\theta_{\rho -}$ and $\theta_{\sigma +}$ 
are not included in this term.  Taking $g_{1\perp}^{++} > 0$ into 
account, we find that the fields $\phi_{\rho -}$ and $\phi_{\sigma +}$ 
are locked at 
$\left\langle\phi_{\rho -}\right\rangle=\frac{\pi}{2} I_1 + \pi I_2$ and 
$\left\langle\phi_{\sigma +}\right\rangle=\frac{\pi}{2} (I_1 + 1)+ \pi I_3$ 
where $I_n$ are integers.

Let us then turn to the $\sigma -$ mode that remains to be studied. 
The effective Hamiltonian of the $\sigma -$ mode takes the form
\begin{eqnarray}
H_{\phi_{\sigma -}} && = \int dx 
\left[K_{\sigma -}\left(\partial_x \theta_{\sigma -} \right)^2 +
K_{\sigma -}^{-1} \left(\partial_x \phi_{\sigma -} \right)^2  \right.
\nonumber \\
&&  \left. + 
g_{\phi} \cos 2\phi_{\sigma -} + g_{\theta} \cos 2 \theta_{\sigma -} 
\right]
\label{eq;hamsm}
\end{eqnarray}
where $g_{\phi} = g_{1\perp}^{+-}-g_{2\perp}^{--}-g_{1\perp}^{++}$ and 
$g_{\theta} = g_{2\perp}^{--}-g_{1\perp}^{--}$. We here adopt a set of 
the variables $(\phi_{\rho -},\phi_{\sigma +}) = (\pi I_1,\pi/2 + \pi I_2)$; 
however, it leads to a physically equivalent result if we chose another set 
$(\phi_{\rho -},\phi_{\sigma +}) = (\pi/2 + \pi I_1,\pi I_2)$.  For the 
Hamiltonian (\ref{eq;hamsm}), there are three RG equations as follows:
\begin{eqnarray}
\frac{dK_{\sigma -}}{d l} =&& y_{\theta}^2 - K_{\sigma -}^2 y_{\sigma}^2 
\\
\frac{dy_{\phi}}{d l} =&& 
\left( 2 - 2 K_{\sigma -} \right)  y_{\phi}
\\
\frac{dy_{\theta}}{d l} =&& 
\left( 2 - 2 K_{\sigma -}^{-1} \right)  y_{\theta}. 
\end{eqnarray}
where $y_\alpha=g_\alpha/(\pi v_F)$ with $\alpha=\theta, \phi$.  
Since $\left| y_{\theta} \right| \gg \left| y_{\phi} \right|$ and 
$K_{\sigma -} > 1$, the parameters are renormalized to 
$y_{\theta} \to \infty$, $y_{\phi} \to 0$, and 
$K_{\sigma -} \to \infty$. As a results, we find that three modes 
are locked as 
$\phi_{\rho -} = \frac{\pi}{2} I_1 + \pi I_2$, 
$\phi_{\sigma +} = \frac{\pi}{2}(I_1 + 1)+ \pi I_3$, and 
$\theta_{\sigma -} = \frac{\pi}{2}(I_1 + 1)+ \pi I_4$.  
Therefore, there is a gap in all the spin excitations.  

\subsection{Order parameters}
\label{OP}

\begin{figure}
\includegraphics[width=5.5cm]{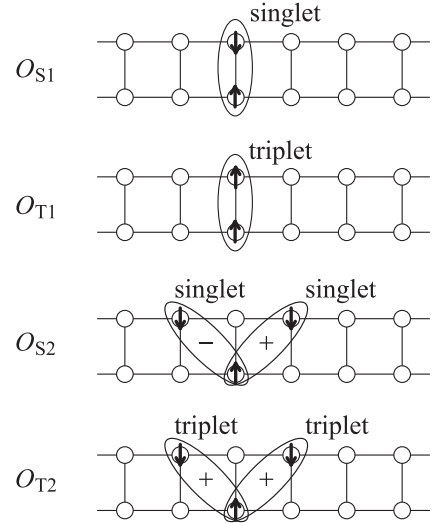}
\caption{\label{fig:sympair} Symmetry of the pairing state 
in the double-chain Hubbard model. The sign of the order 
parameter is also shown.}
\end{figure}

In order to determine the dominant correlation, we introduce the order parameters 
of the superconducting correlations for four kinds of pairing symmetry, which are 
shown in Fig. \ref{fig:sympair}. They are given in terms of the phase fields 
as follows: 
\begin{eqnarray}
O_{S1} = &&
\sum_{p=\pm,\zeta=\pm} 
\psi_{p,\uparrow,\zeta}\psi_{-p,\downarrow,-\zeta} 
\nonumber \\ 
\sim &&
\frac{2}{\pi a} 
e^{i\theta_{\rho +}} \left[
\cos \theta_{\sigma -} \cos \phi_{\sigma +}
\cos \phi_{\rho -} \right. \nonumber \\
&& \left. - i
\sin \theta_{\sigma -} \sin \phi_{\sigma +}
\sin \phi_{\rho -}
\right]
\\
O_{T1} = &&
\sum_{p=\pm,\zeta=\pm} 
\zeta \psi_{p,\uparrow,\zeta}\psi_{-p,\downarrow,-\zeta} 
\nonumber \\ 
\sim &&
\frac{2}{\pi a} 
e^{i\theta_{\rho +}} \left[
i \sin \theta_{\sigma -} \cos \phi_{\sigma +}
\cos \phi_{\rho -} \right. \nonumber \\
&& \left. - 
\cos \theta_{\sigma -} \sin \phi_{\sigma +}
\sin \phi_{\rho -}
\right]
\\
O_{S2} = &&
\sum_{p=\pm,\zeta=\pm} 
\zeta p \psi_{p,\uparrow,\zeta}\psi_{-p,\downarrow,-\zeta} 
\nonumber \\ 
\sim &&
\frac{2i}{\pi a} 
e^{i\theta_{\rho +}} \left[
i \sin \theta_{\sigma -} \sin \phi_{\sigma +}
\cos \phi_{\rho -} \right. \nonumber \\
&& \left. + 
\cos \theta_{\sigma -} \cos \phi_{\sigma +}
\sin \phi_{\rho -}
\right]
\\
O_{T2} = &&
\sum_{p=\pm,\zeta=\pm} 
p \psi_{p,\uparrow,\zeta}\psi_{-p,\downarrow,-\zeta} 
\nonumber \\ 
\sim &&
\frac{2i}{\pi a} 
e^{i\theta_{\rho +}} \left[
\cos \theta_{\sigma -} \sin \phi_{\sigma +}
\cos \phi_{\rho -} \right. \nonumber \\
&& \left. + i
\sin \theta_{\sigma -} \cos \phi_{\sigma +}
\sin \phi_{\rho -}
\right]. 
\end{eqnarray}
One can easily find that the inter-chain diagonal singlet pairing 
state, which corresponds to the order parameter $O_{S2}$, is only of 
the quasi-ordering superconducting instability when $\phi_{\rho -}$ 
and $\theta_{\sigma -}$ are locked.  The asymptotic behavior of the 
S2 correlation function is given by 
\begin{eqnarray}
\left< O_{S2}^{\dagger}(r) O_{S2}(0) \right>
\sim \frac{1}{r^{1/2K_{\rho +}}}.
\label{SS2}
\end{eqnarray}
and the other interchain pairing correlations decay exponentially.  
We also mention that all the intrachain pairing correlations decay 
exponentially due to the locked modes.  
If we assume a S2 ground state $O_{S2}^{\dagger} \left|0\right>$ and 
apply the interacting term of Hamiltonian (\ref{eq:hamint}) to the 
state, we obtain
\begin{eqnarray}
H_I O_{S2}^{\dagger} \left|0\right> = 
(g_{1\perp}^{--}-g_{1\perp}^{+-}-g_{2\perp}^{--}+g_{2\perp}^{+-})
O_{S2}^{\dagger} \left|0\right>.
\end{eqnarray}
Since $g_{1\perp}^{--}<0$, $g_{1\perp}^{+-}>0$, $g_{2\perp}^{--}>0$, 
and $g_{2\perp}^{+-}<0$, we find that the S2 pairing state can gain 
much energy.  

The other possible order is the $4k_{\text F}$ charge-density wave 
(CDW) correlation, of which the order parameter is given by 
\begin{eqnarray}
O_{4k_{\text F}} && = 
\sum_{p,\sigma,\sigma^{\prime}} 
\psi_{p,\sigma,+}^{\dagger} \psi_{p,\sigma^{\prime},-}^{\dagger}
\psi_{-p,\sigma^{\prime},-} \psi_{-p,\sigma,+} \nonumber \\
\sim && 
\frac{1}{a^2 \pi^2} \cos \left(4k_{\text F}x+2\phi_{\rho +} \right)
\left( 
\cos 2\phi_{\sigma +} + \cos 2 \phi_{\sigma -}
\right), \nonumber \\
\end{eqnarray}
where the $\cos 2\phi_{\sigma +}$ term comes from the cases where 
the spins are parallel $\sigma = \sigma^{\prime}$, and the 
$\cos 2\phi_{\sigma -}$ term comes from the cases where the spins 
are antiparallel $\sigma = - \sigma^{\prime}$.  
The latter component $\cos 2\phi_{\sigma -}$ decays exponentially 
because the field $\theta_{\sigma -}$ is locked.  
Thus, the $4k_{\text F}$-CDW correlation shows a power-low behavior 
like 
\begin{eqnarray}
\left<O_{4k_{\text F}}^{\dagger}(r)O_{4k_{\text F}} \right> 
\sim \frac{1}{r^{2K_{\rho +}}}.
\label{4kFCDW}
\end{eqnarray}
By comparing Eq.(\ref{4kFCDW}) with Eq.(\ref{SS2}), we find that 
the $d_{xy}$-wave-like S2 pairing correlation is most dominant 
for $K_{\rho +} > 0.5$, whereas the $4k_{\text F}$-CDW is most 
dominant for $K_{\rho +}< 0.5$.  
This is the same as that of the standard two-band model.~\cite{Schulz96} 

\section{Numerical results}
\label{sec:numerical}

Next, we turn to the intermediate to strong coupling regime.  
We employ the Lanczos and DMRG methods to obtain energies and 
physical quantities in the ground state and low-lying excited 
states.  In order to carry out our calculations, we consider 
$N$ ($=N_\uparrow+N_\downarrow$) electrons in a system with 
length $L$ (containing $2L$ sites).  
The electron density is given by $n=N/L$.  Note that the number 
of electrons must be taken as $N=4l$ with $l$ ($>1$) being an 
integer to maintain the total spin of the unpolarized ground 
state as $S=0$.  

For static quantities, we use the DMRG method with applying 
the open-end boundary conditions for precise calculations.  
We study systems with length up to $L=128$ and keep up to 
$m \approx 2400$ density-matrix eigenstates in the DMRG procedure.  
In this way, the discarded weights are typically of the order 
$10^{-8} \sim 10^{-7}$ and the ground-state energy is obtained 
in the accuracy of $\sim 10^{-7}t$.  All the calculated energies 
are extrapolated to the limit $m \to \infty$.  For dynamical 
quantities, we use the Lanczos method for small clusters with 
the periodic boundary conditions.  The system size is assumed 
as $L=8$, i.e., $8 \times 2$ ladder.  

\subsection{Spin polarization}

\begin{figure}
\includegraphics[width=7.0cm]{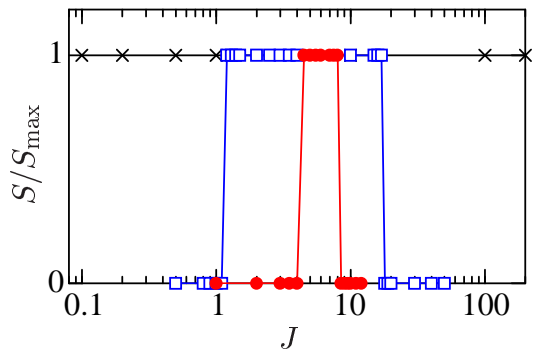}
\caption{(Color online) Calculated values of the total-spin quantum 
number $S$ as a function of $J$ for $U=20$ (circles), $40$ (squares), 
and $\infty$ (crosses).  $L=48$ and $n = 0.5$ are assumed. 
}
\label{fig:fullp}
\end{figure}

Approaching from the strong-coupling regime $U \gg t$, we 
may anticipate the presence of the fully-polarized ferromagnetic 
state.  When the two chains are uncoupled, i.e., $J=0$, 
the state can be interpreted as the Nagaoka state. 
Generally, the appearance of the Nagaoka state is limited 
to large-$U$ and low-$n$ range.  However, the fully-polarized 
region would be spread into smaller-$U$ and higher-$n$ range 
if a finite $J$ is introduced.  This is because the polarized 
electrons can gain the kinetic energy without loss of the exchange 
interaction between the chains.  

Let us start with investigating how the ferromagnetic phase 
appears in the parameter space ($U,J$).  We can find it by 
calculating the expectation value of total-spin operator ${\bf S}$ 
in the ground state, which is defined by
\begin{equation}
\left\langle{\bf S}^2\right\rangle=
\sum_{ij}\left\langle{\bf S}_i\cdot{\bf S}_j\right\rangle
=S(S+1).
\end{equation}
For a fully-polarized state, one will obtain the total-spin quantum 
number $S=S_{\rm max}=N/2$, i.e., $S/S_{\rm max}=1$.  
In Fig.~\ref{fig:fullp}, we show the total spin $S$ normalized with 
respect to $S_{\rm max}$ as a function of $J$ for several values of 
$U$.  The system size and filling are fixed at $L=48$ and $n=0.5$, 
respectively.  We calculate $S$ for systems with length $L=24$, $36$, 
and $48$, and confirm that the size dependence is negligible.  

At $U=20$, with increasing $J$, we find a transition from the 
paramagnetic state to the fully-polarized state at $J_{\rm c1}\sim 4.5$.  
Moreover, we find that the system goes back to the paramagnetic 
state at $J_{\rm c2} \sim 8.5$.  This is because the formation 
of the local spin-singlet bound states gains more energy for very 
large values of $J$ since $J$ increases the gain in the spin-singlet 
binding energy as shown in Sec.\ref{bindene}.  
We also find that the region with the full spin polarization broadens 
with increasing $U$.  In the limit of $U = \infty$, $J_{\rm c1} \to 0$ 
and $J_{\rm c2} \to \infty$.  We note that both of the transitions 
are discontinuous, i.e., of the first-order in the thermodynamic limit.  
Thus, the critical transition points can be determined in the parameter 
space ($U,J$), which will be given as a ground-state phase diagram 
in Sec.\ref{pd}.

\subsection{Excitation gaps}

To ascertain the presence of the lowest excitation gap in the charge, 
spin, and pairing sectors, we calculate the charge gap, spin gap, and 
binding energy in the thermodynamic limit.  We study several chains 
with lengths up to $L=64$ and then perform the finite-size scaling 
analysis based on the size-dependence of each quantity.  

\subsubsection{Charge gap}

\begin{figure}
\includegraphics[width=7.0cm]{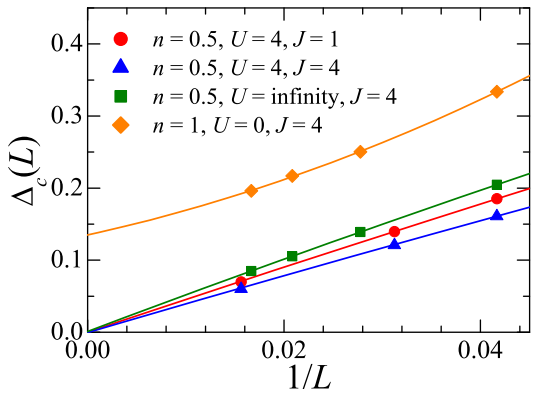}
\caption{(Color online) $\Delta_c (L)$ as a 
function of the inverse system length $1/L$.  Lines are the fits with 
the 2nd-order polynomials. }
\label{fig:chargegap}
\end{figure}

The charge gap $\Delta_c$ is defined by
\begin{eqnarray}
&&\Delta_c = \lim_{L \to \infty} \Delta_c(L), \\
&&\Delta_c(L) = E_L^0 (N_{\uparrow}+1,N_{\downarrow}+1) 
+ E_L^0 (N_{\uparrow}-1,N_{\downarrow}-1) \nonumber \\
&&\hspace*{14mm}- 2 E_L^0 (N_e) \nonumber
\end{eqnarray}
where $E_L^0(N_{\uparrow},N_{\downarrow})$ denotes the ground-state 
energy of a chain of length $L$ with $N_{\uparrow}$ spin-up and 
$N_{\downarrow}$ spin-down electrons.  In Fig.~\ref{fig:chargegap}, 
we plot the charge gap $\Delta_c(L)$ as a function of the inverse 
system length $1/L$ for several parameter sets.  

At half filling ($n=1$), we can easily expect the system to be a 
Mott insulator for $U>0$.  However, we also find the system is insulating 
even for $U=0$ if $J$ is finite.  This is associated with the fact 
that a localized spin-triplet pair is formed on each rung.  
If an electron is add to the rung, an effective on-rung repulsive 
interaction $U_{\rm eff}=U+\frac{J}{2}$ acts.  Thus, we obtain the 
effective Hamiltonian near half filling 
\begin{eqnarray}
H_{\rm eff}^{n \sim 1}&=& -t \sum_{r_x,\sigma}
(c_{r_x,\sigma}^\dag c_{r_x,\sigma} + {\text H.c.}) \nonumber\\
&&+\left( U+\frac{J}{2} \right) \sum_{r_x,\sigma} n_{r_x}^T n_{r_x,\sigma} 
\delta_{\left\langle S_{r_x}^z \right\rangle,\pm\frac{1}{2}}
\end{eqnarray}
where $c_{r_x,\sigma}=c_{r_x,1,\sigma}c_{r_x,2,\sigma}$ is the 
annihilation operator of an electron with spin $\sigma$ at rung $r_x$, 
$n_{r_x,\sigma}=c_{r_x,\sigma}^\dag c_{r_x,\sigma}$ is the number 
operator, $S_{r_x}^z=n_{r_x,\uparrow}-n_{r_x,\downarrow}$ is the 
$z$-component of total spin, and $n_{r_x}^T=T_{r_x}^\dag T_{r_x}$ 
with the spin-triplet operator $T_{r_x}=c_{r_x,1,\sigma}c_{r_x,2,\sigma}$ 
or $\frac{1}{\sqrt{2}}(c_{r_x,1,\sigma}c_{r_x,2,\bar{\sigma}}
+c_{r_x,1,\bar{\sigma}}c_{r_x,2,\sigma}$. 

Let us then turn to the case away from half filling.  There are two 
points to be noted here: one is whether the system is insulating 
at quarter filling ($n=0.5$) as is the case of the Hubbard and 
$t$$-$$J$ ladder; the other is whether the phase separation occurs 
in the fully-polarized state.  As seen in Fig.~\ref{fig:chargegap}, 
all the results for $n=0.5$ are smoothly extrapolated to zero in the 
thermodynamic limit $1/L \to 0$.  At $U=\infty$ and $J=4$, the system 
is in the fully-polarized state.  We therefore conclude that the 
system is metallic in the entire region except $n=1$.

\subsubsection{Spin gap}

\begin{figure}
\includegraphics[width=8.5cm]{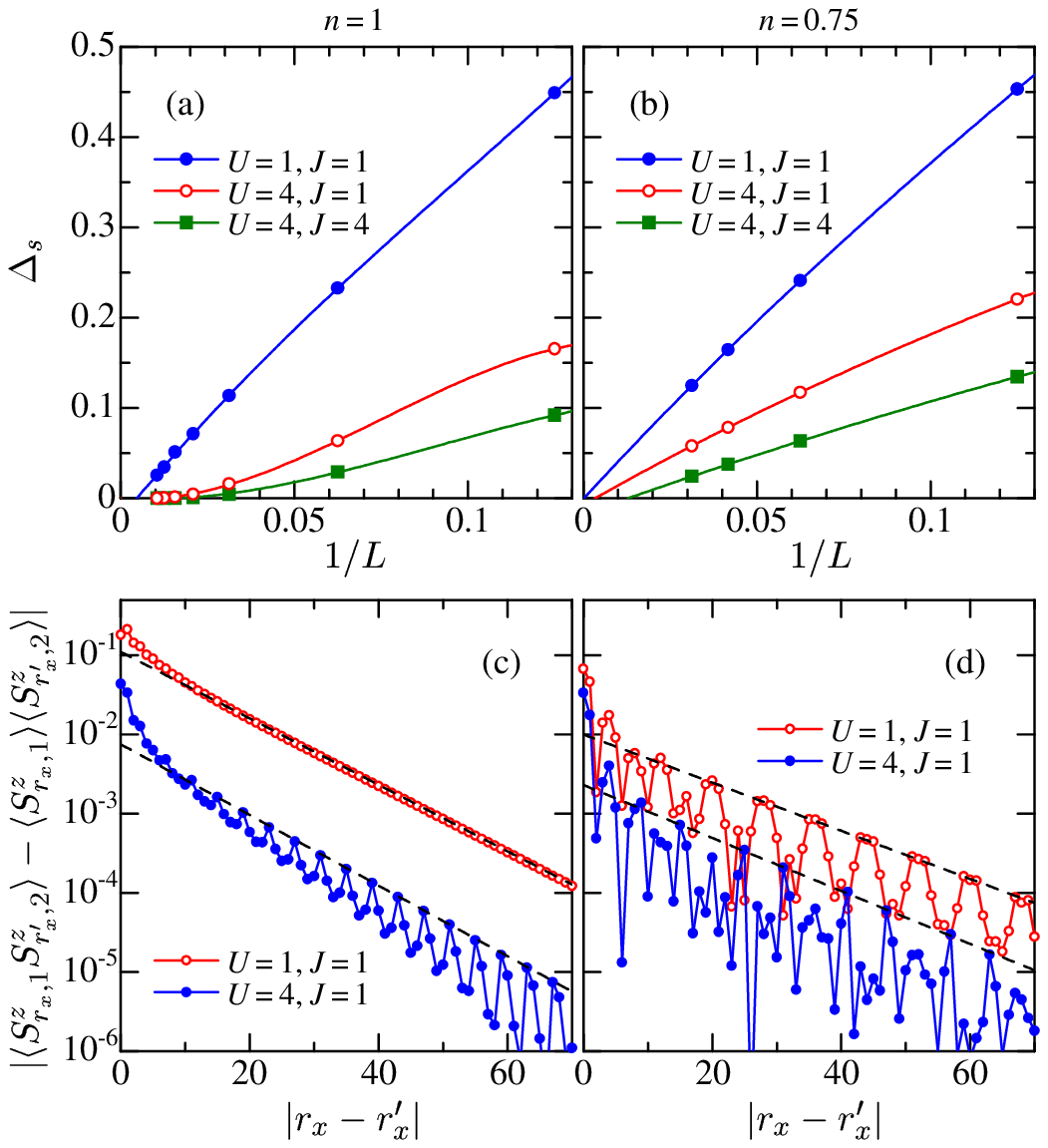}
\caption{(Color online) Upper panels: $\Delta_s(L)$ as a function of 
inverse system size for (a) $n=1$ and (b) $n=0.75$.  Solid lines are 
the polynomial fits to the data for the finite-size scaling analysis.  
Lower panels: Semilogarithmic plot of the magnitude of the spin-spin 
correlation function $S(|r_x-r_x^\prime|,1,2^\prime)$ at $J=1$ for 
(c) $n=1$ and (d) $n=0.75$.  The data are fitted with a function 
$S(|r_x-r_x^\prime|,1,2) \simeq \exp\left(-\frac{|r_x-r_x^\prime|}
{\xi}\right)$.}
\label{fig:spingap} 
\end{figure}

The spin gap $\Delta_s$ is defined by
\begin{eqnarray}
&&\Delta_s = \lim_{L \to \infty} \Delta_s(L), \\
&&\Delta_s(L) = E_L^0 (N_{\uparrow} + 1, N_{\downarrow}-1) 
- E_L^0 (N_{\uparrow}, N_{\downarrow} ).\nonumber
\end{eqnarray}
In Fig.~\ref{fig:spingap}, we plot the spin gap $\Delta_s(L)$ as a 
function of inverse system length $1/L$ at (a) $n=1$ and (b) $n=0.75$.  
For $U=4, n=1$ and $U=1, n=0.75$, values of $\Delta_s(L)$ seem to 
be smoothly extrapolated to zero when $1/L \to 0$, which is in contrast 
to the weak-coupling analysis.  For the other cases, however, the 
extrapolated lines reach zero at finite values of $1/L$.  This is 
unphysical and so it may be a good guess that $\Delta_s(L)$ is fairly 
flat around $1/L=0$, as seen in the results for $U=4, n=1$ and $U=1, 
n=0.75$.  Unfortunately, we cannot treat the large enough systems to 
confirm if this is the case.  Nevertheless, if we assume that the 
$1/L$-dependence of $\Delta_s(L)$ reflects the spinon band structure 
near the Fermi point, the flat behavior of $\Delta_s(L)$ may rather 
imply the presence of a spin gapful state.  Since we cannot detect 
the spin gap less than $\sim 10^{-6}$ within the present calculations, 
the existence of a very small spin gap, i.e., $\Delta(L \to \infty) 
\lesssim 10^{-6}$, is conceivable.  

To determine if the spin gap is present or absent, we also calculate 
the equal-time spin-spin correlation function 
\begin{equation}
S(|r_x-r_x^\prime|,r_y,r_y^\prime)=
\langle S_{r_x,r_y}^z S_{r_x^\prime,r_y^\prime}^z \rangle 
- \langle S_{r_x,r_y}^z \rangle \langle S _{r_x^\prime,r_y^\prime}^z \rangle,
\end{equation}
where $S_{r_x,r_y}^z$ is the z component of the spin operator of an 
electron at site $r_x$ on rung $r_y$.  In Fig.~\ref{fig:spingap}, we 
present a semilogarithmic plot of $|S(|r_x-r_x^\prime|,1,2)|$ as a 
function of the distance $|r_x-r_x^\prime|$ for (c) $n=1$ and 
(d) $n=0.75$.  Note that the long-range behavior of $|S(|r_x-r_x^\prime|,1,1)|$ 
is almost the same as that of $|S(|r_x-r_x^\prime|,1,2)|$.  
The results can be fitted with a function $\exp(-|r_x-r_x^\prime|/\xi)$ 
and thus the exponential decay of the correlation functions is confirmed 
for all the parameter sets.  The correlation lengths are estimated as 
$\xi=10.35 (9.74)$ for $U=1 (4)$ at $n=1$; $\xi=14.29 (12.99)$ for $U=1 (4)$ 
at $n=0.75$.  According to the Tomonaga-Luttinger liquid theory, the 
spin-spin correlation can decay exponentially only when there is a gap 
in all the magnetic excitations.  
Consequently, it would be rather appropriate to conclude that there exists 
quite small spin gap ($\lesssim 10^{-6}$). The correlation lengths seem to 
be much longer than those of other standard spin-gapped systems, 
e.g.,$\xi=3.19$ in the two-leg isotropic Heisenberg system.  However, 
it has been found that in the zigzag Heisenberg chain, the correlation 
lengths increase rapidly with decreasing binding energy.~\cite{White96}  
Thus, the very large values of $\xi$ may reflect an exponentially small 
spin gap.

\subsubsection{Binding energy}
\label{bindene}

\begin{figure}
\includegraphics[width=\hsize]{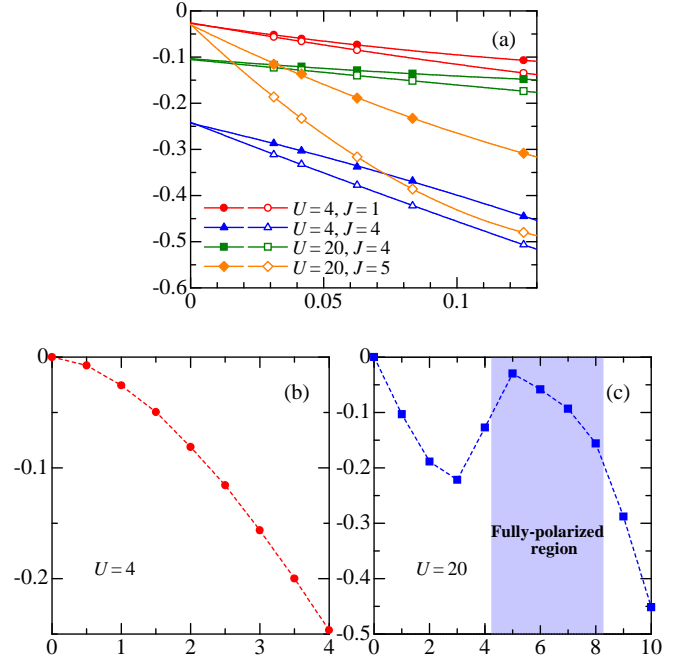}
\caption{\label{fig:binding} (Color online) Upper panel: $\Delta_B^{\pm}(L)$ 
as a function of inverse system length $1/L$ at $n=0.5$.  Solid lines are the 
polynomial fits to the data for finite-size scaling analysis.  Solid (empty) 
symbols denote the binding energy of electrons (holes).  
Lower panels: Values of $\Delta_B^{\pm}(L)$ extrapolated to $1/L \to 0$ 
plotted as a function of $J$ for (b) $U=4$ abd (c) $U=10$.  
}
\end{figure}

The binding energy $\Delta_B^{\pm}$ is defined by
\begin{eqnarray}
&&\Delta_B^{\pm} = \lim_{L \to \infty} \Delta_B^{\pm}(L), \\
&&\Delta_B^{\pm}(L) =  E_L^0 (N_{\uparrow}\pm 1,N_{\downarrow} \pm 1) 
+ E_L^0 (N_{\uparrow}, N_{\downarrow} ) \nonumber \\
&&\hspace*{14mm}-2E_L^0 (N_{\uparrow} \pm 1, N_{\downarrow}) \nonumber
\end{eqnarray}
where the $+$ ($-$) sign corresponds to the binding energy of electrons 
(holes).  
In Fig.~\ref{fig:binding} (a), we plot the binding energy $\Delta_B^{\pm}(L)$ 
as a function of the inverse system length $1/L$ at $n=0.5$ for several 
sets of parameters.  For all the parameter sets, $|\Delta_B^{\pm}(L)|$ is 
found to decrease monotonically as a function of $1/L$, so that we can 
extrapolate $|\Delta_B^{\pm}(L)|$ to the thermodynamic limit systematically.  
We perform a least-squares fit of $|\Delta_B^{\pm}(L)|$ to a polynomial 
in $1/L$ and obtain the extrapolated values.  We note that the binding energies 
of electrons and holes are extrapolated to the same value at $1/L \to 0$.  

In Fig.~\ref{fig:binding} (b), the extrapolated values of $\Delta_B^{\pm}$ 
for $U=4$ as a function of $J$ are shown.  When $U$ is small, the system 
is expected to be in the spin-singlet superconducting phase, as discussed 
in Sec.~\ref{sec:weakcoupling}.  Hence, the binding energy is determined by 
an energy of the spin-singlet bound state.  In analogy with the Haldane gap, 
we expect a scaling $\Delta_B^{\pm} \propto Jt/U^2$ from perturbation assuming 
the double occupancy is sufficiently excluded.  The origin of this energy 
gain is also interpreted as the double exchange interaction in our model.  
For small $J$, this estimation seems not to work well.  However, the double 
occupancy is increasingly excluded with increasing $J$ and thus 
$|\Delta_B^{\pm}(L)|$ increases linearly with $J$ at $J \gtrsim 3$.  
In Fig.~\ref{fig:binding} (c), the extrapolated values of $\Delta_B^{\pm}$ 
for $U=20$ as a function of $J$ are shown.  Since the double occupancy is 
almost excluded, $|\Delta_B^{\pm}|$ increases linearly with $J$ even if $J$ 
is small.  It is notable that the binding energy is strongly suppressed in 
the fully-polarized phase.  This phase is regarded as the spin-triplet 
superconducting one, where the binding energy is determined by an energy of 
the spin-triplet bound state, which is scaled as 
$|\Delta_B^{\pm}| \propto J-J_{\rm c1}$.  
Generally, an energy of the spin-triplet bound state is much lower than that 
of the spin-singlet bound state.~\cite{Yosida66}  

\subsection{Superconducting correlation}

\begin{figure}
\includegraphics[width=8.5cm]{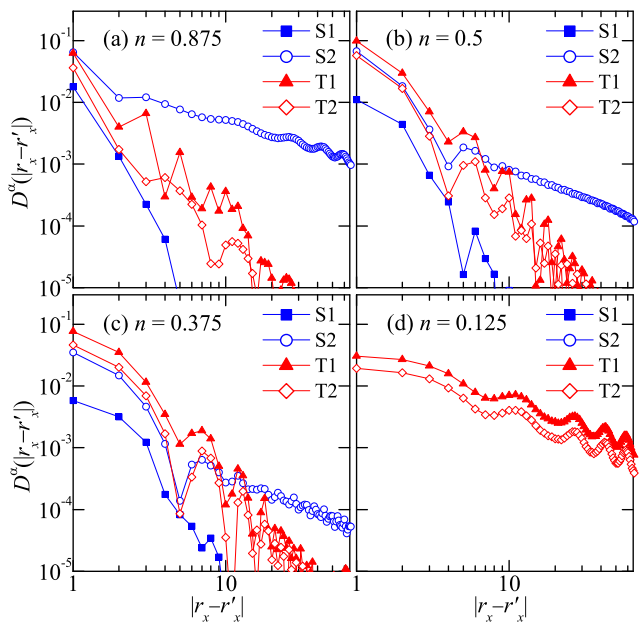}
\caption{\label{fig:corre1} (Color online) Log-log plot of the 
pair-correlation functions 
$D^{\alpha}(|r_x-r_x^\prime|)$ calculated for $L=128$ at 
(a) $n= 0.825$, $U=10$, $J=2$, (b) $n= 0.5$, $U=10$, $J=4$, 
and (c) $n= 0.25$, $U=10$, $J=5$.}
\end{figure}

For our model, we can consider four kinds of superconducting 
correlations as mentioned in Sec.\ref{sec:weakcoupling}.  
In order to estimate them numerically, we define the corresponding 
pair correlation functions as
\begin{eqnarray}
D^{\alpha} (|r_x-r_x^\prime|) = 
\left< 
\Delta_{\alpha}^{\dagger} (r_x)
\Delta_{\alpha}(r_x^\prime)
\right>
\end{eqnarray}
with
\begin{eqnarray}
\Delta_{S1}(r_x) =&&  c_{r_x,1,\uparrow}c_{r_x,2,\downarrow}
- c_{r_x,1,\downarrow}c_{r_x,2,\uparrow} \\
\Delta_{T1}(r_x) =&&  c_{r_x,1,\uparrow}c_{r_x,2,\downarrow}
+ c_{r_x,1,\downarrow}c_{r_x,2,\uparrow} \\
\Delta_{S2}(r_x) =&&  c_{r_x,1,\uparrow}c_{r_x+1,2,\downarrow}
- c_{r_x,1,\downarrow}c_{r_x+1,2,\uparrow} \\
\Delta_{T2}(r_x) =&&  c_{r_x,1,\uparrow}c_{r_x+1,2,\downarrow}
+ c_{r_x,1,\downarrow}c_{r_x+1,2,\uparrow}, 
\end{eqnarray}
which are calculated by the DMRG method.  
The calculated results for three sets of parameters are shown in 
Fig.~\ref{fig:corre1}.  For $n = 0.825$, $U=10$, and $J=4$ 
[Fig.~\ref{fig:corre1}(a)], the S2 pairing correlation is 
clearly the most dominant one, which shows a power-law length 
dependence.  The ratio of the decay is estimated to be 
$\sim r^{-0.7}$, which leads to $K_\rho^+=0.714$.  This is 
consistent with the bosonization result for the spin gapful 
state.  At $n = 0.5$, $U = 10$, $J = 4$ [Fig.~\ref{fig:corre1}(b)], 
the system has a $S=0$ ground state but is somewhat closer to 
the fully-polarized ferromagnetic phase.  The S2 pairing correlation 
is still the most dominant one but the T1 pairing correlation 
becomes much enhanced at a short distance $|r_x-r_x^\prime| \lesssim 10$.  
This is because the formation of a local spin-triplet bound state 
on a rung can gain some energies.  We note that the T2 pairing 
correlation is also enhanced, reflecting a tendency to the 
fully-polarized ferromagnetic state.  The decay ratio of the S2 
correlation is $\sim r^{-0.8}$, which leads to $K_\rho^+=0.625$.  
This value agrees well with our direct estimation of $K_\rho^+$ 
(See Sec.\ref{TL}).  If the system further approaches the 
fully-polarized ferromagnetic phase [Fig.~\ref{fig:corre1}(c)], 
the change in the correlation functions at short ranges becomes 
more prominent.  Surprisingly, we find that the decay lengths of 
all the correlations are almost unchanged as far as $J$ is fixed 
in the $S=0$ ground state.  In the fully-polarized ferromagnetic 
phase [Fig.~\ref{fig:corre1}(d)], the T1 pairing correlation is 
the most dominant one and only the S2 pairing correlation is 
competing.  The decay ratio of both correlation functions is 
$\sim r^{-1}$, which leads to $K_\rho^+=0.5$.  

\subsection{Anomalous Green function}

\begin{figure}
\includegraphics[width=8.5cm]{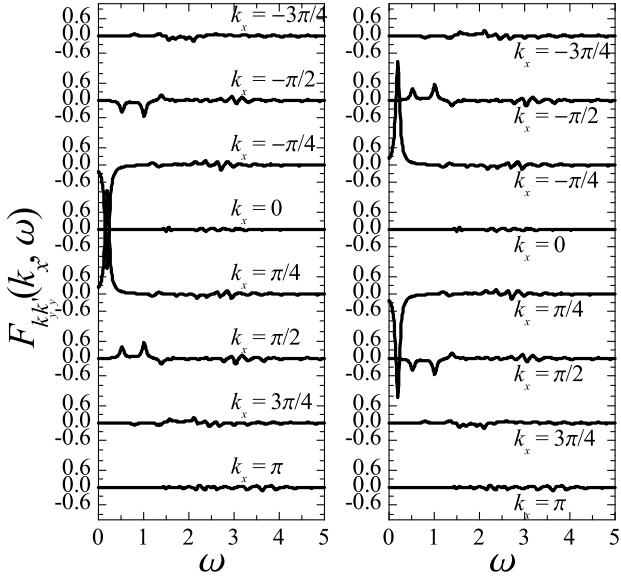}
\caption{\label{fig:agr} Left panel: Anomalous Green's function at 
$k_y = 0$, $k_y^{\prime} = \pi$, $n = N_e/N = 5/8$, $U = 10$, 
$J = 2$, and $N = 16$.  
Right panel: Same as the left panel but at $k_y = \pi$, 
$k_y^{\prime} = 0$.}
\end{figure}

Let us now determine the pairing symmetry in the S2 superconducting 
state.  To this end, we calculate the one-particle anomalous Green 
function,~\cite{Ohta94} which exhibits the excitations of the Bogoliubov 
quasiparticles in the BCS theory.  Therefore, we can see how the 
nodes appear in the superconducting gap function.  
The anomalous Green function is defined by 
\begin{eqnarray}
&& G_{k_y,k_y^{\prime}} \left(q_x,z \right) = \nonumber \\
&& \left< \psi_0^{N_e-2} \right| 
c_{q_x,k_y,\uparrow} \frac{1}{z-H + E_0} 
c_{-q_x,k_y^{\prime},\downarrow} \left| \psi_0^{N_e} \right> 
\nonumber \\
\label{eq:agr}
\end{eqnarray}
where $\left|\psi_0^{N_e} \right>$ denote the wavefunction of 
the ground state with $N_e/2$ up-spin and $N_e/2$ down-spin electrons, 
and $E_0$ is chosen as the average value of 
$E_L^0 (N_{\uparrow}-1,N_{\downarrow}-1)$ and 
$E_L^0 (N_{\uparrow},N_{\downarrow})$.  
We then estimate the spectral function as 
\begin{eqnarray}
F_{k_y,k_y^{\prime}}(k_x)= -\frac{1}{\pi} 
\mathrm{Im} G_{k_y,k_y^{\prime}}(q_x,\omega+i\eta)
\end{eqnarray}
with $\eta = 0^+$ and its frequency integral as
\begin{eqnarray}
F_{k_y,k_y^{\prime}}(k_x)=\left< \psi_0^{N_e-2} \right| c_{k_x,k_y,\uparrow}
c_{-k_x,k_y^{\prime},\downarrow} \left| \psi_0^{N_e} \right>.
\end{eqnarray}
We calculate the anomalous Green function (\ref{eq:agr}) for the 
ladder with length $L=8$ by using the Lanczos method.  We should 
note that the BCS theory is not readily applicable for (quasi) 
one-dimensional system; actually, $F_{k_y,k_y^{\prime}}(q_x)$ is 
not long-range ordered with a logarithmic decay as a function of $L$.  
However, apart from this prefactor, we can naively expect that 
$F_{k_y,k_y^{\prime}}(q_x)$ calculated on finite-size systems 
should provide information on the pairing symmetry at intermediate 
distances.  

The calculated results for $F_{k_y,k_y^{\prime}}(k_x,\omega)$ are 
shown in Fig.~\ref{fig:agr}.  We can observe pronounced low-energy 
peaks at $|k_x|=\frac{\pi}{4}$ for all $k_y$ and $k_y^\prime$, 
which are the nearest to the Fermi momenta $k_{\rm F}(=\frac{5}{16}\pi)$, 
and much smaller peaks at higher energies for other momenta.  
We also find that the spectral weight vanishes at $q_x = 0$, 
$q_x = \pi$, and $k_y = k_y^{\prime}$.  The weights of the peaks 
appear to be similar to the BCS form of the condensation amplitude, 
which has a maximum value at the Fermi momenta.  We may thus assume 
that the superconducting ground state in strongly correlated electron 
systems can be characterized by a gap function, which is directly 
proportional to the frequency-integrated function 
$F_{k_y,k_y^{\prime}}(q_x)$.~\cite{Ohta94,Poilblanc03} 
The function $F_{k_y,k_y^{\prime}}(k_x)$ obeys the relation 
$F_{k_y,k_y^{\prime}}(k_x) = -F_{k_y^{\prime},k_y}(k_x)$ and 
$F_{k_y,k_y^{\prime}}(k_x) = - F_{k_y,k_y^{\prime}}(-k_x)$.  
This indicates the formation of the $d_{xy}$-wave like pairing state 
in our system, which is consistent with the weak-coupling results 
shown in Sec.\ref{OP}. 

\subsection{Tomonaga-Luttinger liquid parameter}
\label{TL}

\begin{figure}
\includegraphics[width=6.0cm]{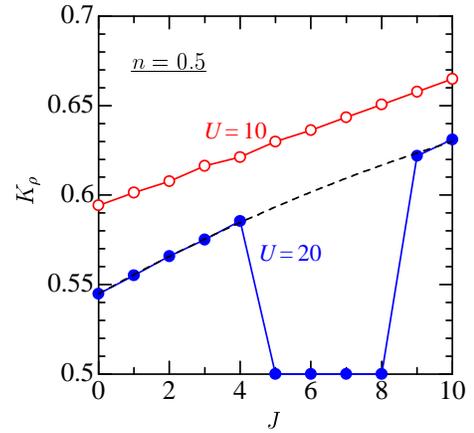}
\caption{\label{fig:krho} (Color online) DMRG results for 
$K_{\rho +}$ as a function of $J$.  The results for $U = 10$ 
and $20$ at $n=0.5$ are shown.  Dotted line is guide for eyes.}
\end{figure}

For the calculation of the TL parameter, we use a recently 
proposed method of the DMRG technique.~\cite{Ejima05}  
As mentioned in Sec.~\ref{OP}, the TL parameter $K_\rho^+$ 
determines the long-range decays of the $d_{xy}$-like (S2) 
pairing and $4k_F$-CDW correlations in the metallic TL-liquid 
ground state, whereas the parameter $K_\rho^-$ is scaled as 
$K_\rho^- \to 0$.  
For the double-chain model, we define the TL parameter as 
\begin{eqnarray}
K_{\rho \pm} =&& \frac{\pi}{2} \lim_{q \to 0} 
\frac{\partial N_{\pm} (q)}{\partial q} 
\end{eqnarray}
with
\begin{eqnarray}
N_{\pm} (q) =&& \frac{1}{L} \sum_{r_x,r_x^{\prime}} 
e^{iq (r_x - r_x^{\prime})}
\left< n_{r_x}^{\pm}n_{r_x^{\prime}}^{\pm} \right>.
\end{eqnarray}
where $n_{r_x}^{\pm} = n_{r_x,1}\pm n_{r_x,2}$.  
In Fig.~\ref{fig:krho}, we show the calculated results for 
$K_{\rho +}$ as a function of $J$ for $U = 10$ and $20$ 
at $n=0.5$.  For $U=10$, $K_\rho^+$ increases monotonously 
with increasing $J$, which is consistent with the monotonous 
increase in the binding energy with respect to $J$ in the 
strong-coupling regime.  
For $U=20$, the behaviors for $J \lesssim 4$ and $J \gtrsim 9$ 
are quite similar to those for $U=10$, although the values 
are somewhat smaller due to the effects of the Umklapp scattering.  
However, we estimate the TL parameter as $K_\rho^+=0.5$ for 
$5 \lesssim J \lesssim 8$.  This regime corresponds to the 
fully-polarized ferromagnetic phase and the TL parameter should 
be the same as that of a spinless fermion system.  
We thus find $K_\rho>1/2$ in the entire region except in the 
fully-polarized ferromagnetic phase and we confirm that the 
S2-type superconducting correlation is most dominant.  

\subsection{Phase diagram}
\label{pd}

\begin{figure}
\includegraphics[width=7.5cm]{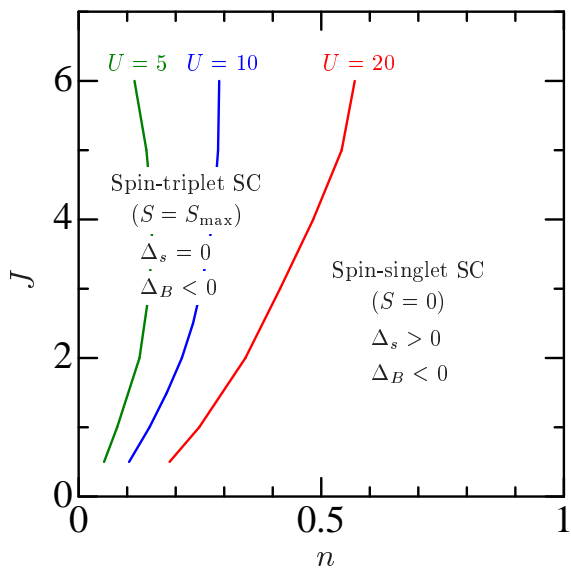}
\caption{\label{fig:PD} (Color online) Phase diagram of our 
double-chain Hubbard model obtained from the DRMG calculations.  
The phase boundary between the fully-polarized ferromagnetic 
phase and $d_{xy}$-like superconducting (SC) phase is shown 
for $U=5$, $10$, and $20$. }
\end{figure}

In Fig.~\ref{fig:PD}, we show the phase diagram of our model 
in the parameter space $(n, J)$ for $U=5$, $10$, and $20$.  
The phase boundary is determined by the calculated results 
for the total-spin quantum number.  At higher electron 
concentrations from the phase boundary, the system is 
characterized as the S2-type spin-singlet superconducting state, 
and at lower concentrations from the phase boundary, the system 
is characterized as the fully-polarized ferromagnetic state and 
simultaneously as the T1-type spin-triplet superconducting state.  
We note that the fully-polarized ferromagnetic phase is spread to 
the higher concentration range with increasing $U$, and it 
occupies the entire parameter region ($n<1, J>0$) at $U=\infty$.  
This result is connected to the Nagaoka state\cite{Nagaoka66} at 
$U \to \infty$.  

\section{Summary and discussion}
\label{sec:summary}

We have studied the ground-state properties of the double-chain 
Hubbard model coupled with ferromagnetic exchange interaction by 
using the weak-coupling theory, DMRG technique, and the Lanczos method.  
We have thereby determined the ground-state phase diagram in the 
parameter space of the ferromagnetic exchange interaction and 
band filling.  We have found that, in high electron density regime, 
the spin gap opens and the spin-singlet $d_{xy}$-wave-like pairing 
correlation is most dominant, and in low electron density regime, 
the fully-polarized ferromagnetic state is stabilized and simultaneously 
the spin-triplet $p_{y}$-wave-like pairing correlation becomes 
most dominant.  

Here, let us make some comment on what happens if some additional 
terms are introduced to our model.  First, we consider the case where 
the hopping integral between the two chains $t_\perp$ is added.  
In this case, the couplings $g_{1\perp}^{--}$ and $g_{2\perp}^{+-}$ 
become irrelevant, which leads to a collapse of the pairing state 
between different bands, i.e., $k_y = 0$ and $\pi$.  Thus, the 
spin-singlet superconductivity is suppressed.  Since the term $t_\perp$ 
induces the antiferromagnetic interaction on each rung, the spin-triplet 
superconductivity may also be suppressed.  Next, we consider the case 
where the intersite Coulomb interaction between the two chains $V_\perp$ 
is added.  In this case, we can easily imagine that the spin-triplet 
superconductivity is strongly suppressed because the $d_y$-wave-like 
pairing state is formed on rung.  The double exchange interaction 
for the $d_{xy}$-wave-like pairing state is also suppressed.  

Finally, let us discuss possible relationship between the present 
model and the model of two Hubbard chains coupled with zigzag bonds 
where the spin-triplet superconductivity has been shown to 
occur.\cite{Ohta05,Nishimoto08a,Nishimoto08b}  The two models in 
the strong-coupling regime have the following features in common: 
(i) the superconductivity occurs near the region of ferromagnetism, 
(ii) there is a competition between the spin-singlet and spin-triplet 
pairings, and (iii) the spin-gap is quite small or vanishes.  
These situations may well suggest that the spin-triplet pairing 
could be dominant in the present model near the phase boundary 
between the spin-singlet superconductivity and ferromagnetism.  
So far, we have not found any indications that the spin-triplet 
pairing becomes more dominant than the spin-singlet pairing unless 
the ground state is spin polarized.  
We hope that this point will be clarified with more elaborate 
calculations done in future.  

\begin{acknowledgments}
We thank S. Ejima, H. Matsueda, K. Sano, and H. Yoshioka for 
useful discussions. 
This work was supported in part by Grants-in-Aid for 
Scientific Research (Nos. 18028008, 18043006, 18540338, and 19014004) 
from the Ministry of Education, Science, Sports and Culture of Japan.  
A part of computations was carried out at the Research Center for 
Computational Science, Okazaki Research Facilities, and 
the Institute for Solid State Physics, University of Tokyo.  
\end{acknowledgments}

\end{document}